%% file: arxiv.tex
\documentclass[%
 amsmath,amssymb,
 aps,
 floatfix,
 twocolumn
]{revtex4-2}

\include{inc/preamble}

\begin{document}

\include{inc/preface}
\maketitle

\section{Introduction}

Computational materials and molecular modelling have proved to be an invaluable tool in predicting new materials and processes and interpreting experimental phenomena on the microscopic level.
The predictive performance of atomistic simulations strongly depends on the accuracy of the employed atomic interaction model, of which those based on solving the Schr\"odinger equation are generally regarded as the most reliable.
However, many problems of interest remain intractable, even when using approximate solutions of the quantum mechanical problem, such as \ac{DFT}, due to the high computational cost and its scaling with respect to system size.
While interatomic potentials based on simple analytical forms open up access to significantly larger system sizes and longer simulation times, their parameterisation is often insufficiently accurate for predictive modelling.
Surrogate models for quantum mechanical calculations, based on highly flexible functional forms provided by machine learning methods which are fitted using high-quality \abinitio{} reference data emerged in the last two decades~\cite{Behler.2007,Shapeev.2016,Thompson.2015,Drautz.2019,Artrith.2016,Bartok.2010,Schutt.2018}.
These \acp{MLIP} reproduce the \abinitio{} potential energy surface to a high accuracy in a computationally efficient way, allowing access to large time and length scale simulations~\cite{Deringer.2020}.

In this work, we focus on the fitting aspect of \acp{MLIP}, \emph{i.e.} the process that determines the model parameters based on a set of reference data points.
Even though fitting is typically a one-off operation, and its computational cost leaves the cost of a subsequent simulation using the \ac{MLIP} largely or completely unaffected, it can use significant resources and can be a limiting factor in applying ever increasing data bases or exploring the space of model hyperparameters.
Depending on the regression method, some \acp{MLIP} are based on solving a linear system to obtain the model weights.
Here we present a set of application principles that can be used to distribute the workload among multiple processes when fitting such models, allowing efficient utilisation of massively parallel computer resources.
We have implemented these in the \ac{GAP} framework and demonstrated excellent scaling in both memory and computational efficiency up to thousands of cores.
We note that similar efforts have been made to parallelise the \code{fitSNAP} code used for fitting \ac{SNAP}~\cite{Thompson.2015} linear models~\cite{fitSnap}.
Other \ac{MLIP} implementations which are based on kernel methods~\cite{Lilienfeld.2015} or linear regression, such as  the \ac{LML}~\cite{Goryaeva.2019} and \ac{ACE}~\cite{Drautz.2019} approaches, would also benefit from similar developments.

%Solid-state, DFT, configurations, expensive, machine learning, Gaussian process (GP), Gaussian Approximation Potential (GAP)\cite{GAP}, sparse/basis set

\section{Theory}\label{sec:theory}

We provide a brief overview of the \ac{GAP} framework, and for more details we refer the reader to a more comprehensive review~\cite{Deringer.2021} of the method.
\Ac{GAP} is a Bayesian regression method that aims to create surrogate models for the quantum mechanical interactions by using a reference database consisting of atomic configurations and associated microscopic properties obtained from \abinitio{} calculations.
Strategies to create such databases are discussed elsewhere~\cite{Li2015,Vandermause2020}; here we start from a set of configurations, each consisting of Cartesian coordinates with the corresponding atomic species information, and for \abinitio{} data we use total energies, forces and virial stress components.
As Cartesian coordinates do not transform in an invariant fashion when applying energy-conserving symmetry operations to the atomic positions, such as rotations, translations, inversion and permutation of the indices of identical atomic species, it is beneficial to first transform the Cartesian coordinates to descriptors, which form the input vectors $\bx_i$ of length $d$ to the regression method.

In general, \ac{GAP} approximates the total energy of a configuration $A$ \added{as a sum of local terms
\begin{equation}
    E_A = \sum_{i=1}^{N_A} \varepsilon (\bx_i)
    \textrm{,}
\end{equation}
where the local energy term $\varepsilon$ is} in the form of a sparse \ac{GP}~\cite{QuinoneroCandela:2005wp,Snelson:2006vi}
\added{
\begin{equation}
    \varepsilon(\mathbf{x}_i) =  \sum_j^M c_j k(\bx_i,\bx_j)
\end{equation}}
\noindent where \deleted{the first sum includes all descriptor vectors in configuration $A$, and} the \deleted{second} sum is over a set of representative descriptor vectors, or sparse points $M$.
The kernel function $k(\bx_i,\bx_j)$ evaluates the similarity of descriptors $\bx_i$ and $\bx_j$, and $c_j$ are the regression weights that need to be fitted such that predicted properties match the \abinitio{} values as closely as possible.
Forces and virial stress components can be obtained by differentiating this expression with respect to atomic coordinates or lattice deformations, which is a trivial, but rather tedious operation and we omit it from here for brevity.
Denoting the $N$ \abinitio{} reference properties by $\by$ and predicted properties by $\tilde{\by}$, we formulate the regression problem as minimising the loss function
\begin{equation}
    \mathcal{L} = (\by - \tilde{\by})^T \bSigma^{-1} (\by - \tilde{\by}) + \bc^T \bK_{MM} \bc
    \label{eq:loss}
\end{equation}
with respect to the weights $\bc$.
\added{We note that elements of $\by$ may be total energies or their derivatives, such as force and stress components, but here we only review the case for total energies.}
The matrix $\bSigma$ is diagonal and its elements are inversely related to the importance of each data point.
While the first term is responsible for achieving a close fit to the data points, the second term is controlling overfitting via a Tikhonov regularising expression, which forces the elements of $\bc$ to remain small.
The elements of the matrix $\bK_{MM}$ are built from the kernel function values $K_{ij} = k (\bx_i, \bx_j)$ between the sparse point set $\{\bx_i\}_{i=1}^M$ where we typically use $M \ll N$.
The minimum of the loss function in \cref{eq:loss} can be determined analytically \replaced{as a solution of the linear system}{, and the result is}
\begin{equation}
    (\mathbf{K}_{MM} + 
    \mathbf{K}_{MN} \bSigma^{-1} \mathbf{K}_{NM}) \mathbf{c}  =  \mathbf{K}_{MN} \bSigma^{-1} \by
    \label{eq:GP_sparse}
\end{equation}
\added{for $\bc$, or equivalently}
\begin{equation}
    \mathbf{c}  = (\mathbf{K}_{MM} + 
    \mathbf{K}_{MN} \bSigma^{-1} \mathbf{K}_{NM})^{-1} \mathbf{K}_{MN} \bSigma^{-1} \by
    \textrm{.}
    \label{eq:GP_sparse_inv}
\end{equation}
The elements of $\bK_{MN}$ are given by
\begin{equation}
    K_{ij} = \sum_{\alpha = 1}^{N_j} k(\bx_i,\bx_\alpha)
\end{equation}
where $\bx_i$ is a descriptor vector from the sparse set, and $j$ denotes a target total energy and the sum includes all  descriptors \added{that correspond to local energy terms} \deleted{that} \replaced{contributing}{contribute} to $y_j$.
For convenience, we use the notation $\bK_{MN}^T \equiv \bK_{NM}$.
Elements of $\bK_{MN}$ corresponding to derivative observations are calculated similarly, using the appropriate gradients of the kernel function $k$, for which further details may be found in the review article by Deringer \etal{}~\cite{Deringer.2021}.

The complexity of solving \cref{eq:GP_sparse} scales with $\mathcal{O}(M^2 N)$, which is significantly more favourable than the $\mathcal{O}(N^3)$ scaling of a full \ac{GP} implementation.
However, Foster \etal{} have shown~\cite{Foster:2009wy} that the solution may lead to numerically unstable results at large data sets, i.e. uncertainties in the input lead to disproportionate errors in the output. 
Following their suggestions, we first define the $(N+M)\times M$ matrix 
\begin{equation}
    \mathbf{A} = \begin{bmatrix}
    \bSigma^{-\sfrac{1}{2}}\mathbf{K}_{NM} \\
    \mathbf{L}_{MM}^T
    \end{bmatrix}
\end{equation}
where the lower triangular matrix $\mathbf{L}_{MM}$ is the result of the Cholesky decomposition of $\mathbf{K}_{MM}$ such that $\mathbf{K}_{MM} = \mathbf{L}_{MM}\mathbf{L}_{MM}^T$.
Introducing $\bb$ by padding the vector of target properties $\by$ by an $M$-long vector of zeros
\begin{equation}
    \mathbf{b} = \begin{bmatrix}
        \by \\
        \mathbf{0}
    \end{bmatrix}
\end{equation}
we rewrite \cref{eq:GP_sparse} as the solution of the least-squares problem
\begin{equation}
    \min_\mathbf{c} (\mathbf{A}\bc - \mathbf{b})^T (\mathbf{A}\bc - \mathbf{b})
\end{equation}
that leads to the solution in the form of
\begin{equation}
    \bc = (\mathbf{A}^T \mathbf{A})^{-1} \mathbf{A}^T \bb
    \textrm{.}
    \label{eq:sparseAsolution}
\end{equation}
A numerically stable solution can be obtained by first carrying out a QR factorisation of $\bA = \bQ \bR$ where $\bQ$ is orthogonal, namely, it is formed by \emph{orthonormal} column vectors:
\begin{equation}
    \bQ^T \bQ = \bI \textrm{,}
\end{equation}
while $\bR$ is an upper triangular matrix.
Substituting the factorised form of $\bA$ into \cref{eq:sparseAsolution} results in
\begin{equation}
    \bc = (\bR^T \bQ^T \bQ \bR)^{-1} \bR^T \bQ^T \bb = \bR^{-1} \bQ^T \bb
    \textrm{.}
\end{equation}
The computational complexity of creating $\bA$ is determined by the cost of creating its two constituent blocks.
The calculation of the upper block scales as $\mathcal{O}(MN)$, due to $\bSigma$ being diagonal, while the Cholesky factorisation resulting in the lower block scales as $\mathcal{O}(M^3)$, resulting in an overall scaling $\mathcal{O}(MN)$, as $N \gg M$.
The QR factorisation of $\bA$ requires $\mathcal{O}(M^2N)$ floating point operations, hence dominating the overall cost of evaluating \cref{eq:sparseAsolution}.
We note that multiplying by $\bR^{-1}$ can be implemented as a series of back substitution operations, due to the upper triangular matrix form of $\bR$.

\section{Implementation}

The workflow of obtaining the sparse or representative points and associated vector of weights $\bc$ from a set of reference \abinitio{} configurations is implemented in the \gapfit{} program, and distributed as part of the software package \Ac{QUIP}, which is a Fortran package implementing atomistic simulation tools, including low-level functions to manipulate atomic configurations, a selection of interatomic potentials, tight-binding models and the \ac{GAP} framework.
The source code is publicly available on Github~\cite{QUIPGAPGithub}.

\subsection{Program structure}

\begin{figure}
    \includegraphics[width=0.7\columnwidth]{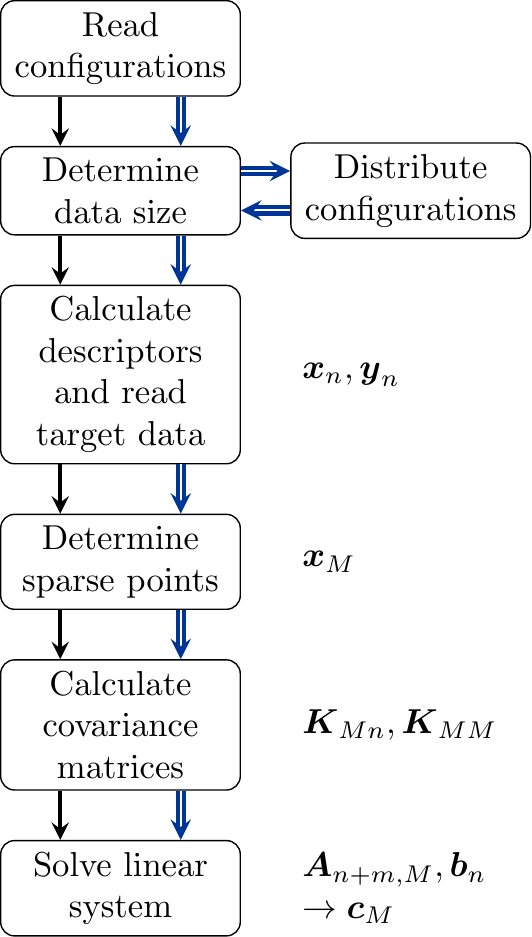}
    \centering
    \caption{Schema of \gapfit{} using serial/thread-parallel (black arrows) and data-parallel (blue arrows) execution code paths.}
    \label{fig:structure}
\end{figure}

The \gapfit{} program is controlled via a set of command line arguments consisting of key-value pairs, which can also be passed as a configuration file.
The program also requires a set of reference configurations in the extended XYZ format~\cite{XYZ}, containing any combination of total energies, forces and virial stresses, and optionally, the definition of a baseline potential model.
The major steps of the fitting process are outlined in \cref{fig:structure}. After initialisation and reading of the command line arguments, the training structures are parsed for the number of target properties: total energies, forces and virial stress components, to determine the value of $N$.
Based on the descriptors, the amount of storage space needed for the descriptor arrays $\bx$ and their derivatives $\bx'$ with respect to Cartesian coordinates are calculated and then allocated.
%according to these numbers, and filled when processing of the input fully.

From the descriptor vectors, $M$ are chosen as a representative (sparse) set.
The procedure for choosing can be controlled by command line arguments, including selecting a random subset, clustering and CUR-based approaches~\cite{Mahoney.2009}.
It is also possible to provide the chosen representative points via files, an option we make use of for the parallel version (see \cref{sec:parallel}).

After setting the sparse points, the covariance matrices $\bK_{MN}$ and $\bK_{MM}$ are calculated.
From these, matrix $\bA$ is constructed and \cref{eq:sparseAsolution} is solved via QR decomposition using linear algebra routines, using \Ac{LAPACK} for single node applications.

\begin{figure}
    \centering
    \includegraphics[width=\columnwidth]{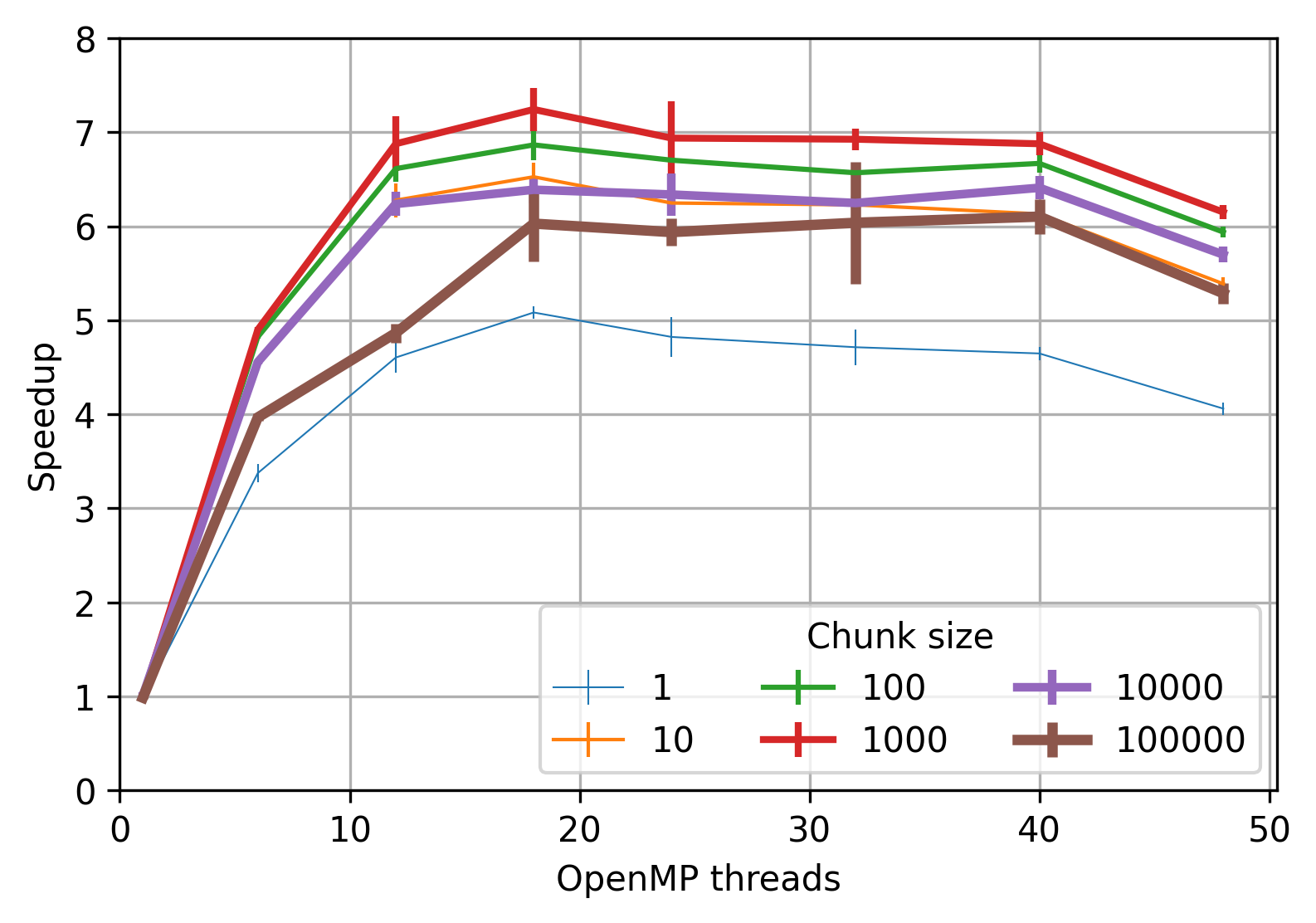}
    \caption{\added{Scaling of computing time of a non-MPI \gapfit{} calculation with the number of \ac{OpenMP} threads for different chunk sizes on a node of the Avon cluster\cite{Avon}. The reference time is \SI{659}{s}, error bars depict twice the standard error.}}
    \label{fig:openmp_scaling}
\end{figure}

The intermediate processing, such as the computation of the elements of covariance matrices had already been augmented by \Ac{OpenMP} directives along the target data dimension $N$, which leads to a thread-based parallelisation on a single process.
This, however, restricts the program to the memory and processing resources of a single node, and performance is further limited by the fact that the speed a computational core can access an arbitrary memory region is inhomogeneous due to the physical layout of the memory. 
That results in a decrease of efficiency when additional threads are employed, leading to a degradation of performance which prevents full utilisation of all available cores in a node.
We present the parallel scalability of a test problem in \cref{fig:openmp_scaling}, where we varied the size of contiguous subsets of \ac{OpenMP} loops, referred to as \emph{chunks}.

As an example of the limitations imposed by the  OpenMP implementation of \texttt{gap\_fit}, the practical problem of fitting a \ac{GAP} for carbon~\cite{Rowe.2020} --- one of the largest single-element training datasets assembled to date --- took more than 6 days on a single node and required more than \SI{1}{TB} memory to accommodate the design and covariance matrices~\cite{gc:private}.
This restricted the ability of practitioners to build complete training sets or to experiment with choices of hyperparameters.

\subsection{Multi-process parallelisation}
\label{sec:parallel}

To go beyond the limitations posed by the shared memory requirement, poor parallel performance, and specialist hardware, we propose a multi-process framework with data distribution and inter-node communication.
We have established in \cref{sec:theory} that both the memory requirement and the computational effort scale linearly with the number of target properties $N$, therefore it is convenient to distribute memory and tasks along this dimension of the problem.

The two most memory intensive operations are the calculation of the descriptor vectors together with their gradients, and the covariance matrices. The ratio of these depends strongly on the particulars of the fitting problem, \replaced{especially}{in particular} the dimensionality $d$ of descriptor vectors and the number of sparse points $M$. 
In our parallel scheme, we distribute atomic configurations across independent processes, such that the number of target properties are as even as possible.
We note that the size of individual atomic configurations may be highly inhomogeneous, therefore the number of forces per configuration can vary substantially across the database, necessitating an extra step that determines the optimal spread of data.
We have employed a greedy algorithm that first collects configurations in a list and sorts them by descending number of target properties. We then assign the largest (by target property) unassigned configuration to the process which currently has the least total number of target properties. This process repeats until the list is exhausted.

With the configurations allotted to \ac{MPI} processes, the descriptor calculations may proceed locally, and once completed, individual portions of $\bK_{MN}$, denoted as $\bK_{Mn}$ can be evaluated.
For this, local copies of the sparse set of $M$ descriptor values need to be present locally, the particulars of which we discuss later in \cref{sec:sparse}.
The na\"ive solution of the linear system represented by \cref{eq:GP_sparse} may be adapted trivially to a distributed $\bK_{MN}$: the terms $\mathbf{K}_{MN} \bSigma^{-1} \mathbf{K}_{NM}$ and  $\mathbf{K}_{MN} \bSigma^{-1} \by$ can be calculated locally and reduced across processes as
\begin{equation}
    \mathbf{K}_{MN} \bSigma^{-1} \mathbf{K}_{NM} = \sum_{n \in N} \mathbf{K}_{Mn} \bSigma_n^{-1} \mathbf{K}_{nM}
\end{equation}
and
\begin{equation}
   \mathbf{K}_{MN} \bSigma^{-1} \by = \sum_{n \in N} \mathbf{K}_{Mn} \bSigma_n^{-1} \by_n
\end{equation}
where we denote distributed blocks of $\bSigma$ and $\by$ by $\bSigma_n$ and $\by_n$, respectively.
The rest of the calculation only involves matrices up the size of $M\times M$.
However, the direct solution, as described in \cref{sec:theory} is numerically unstable, therefore we need to adapt the solution based on the QR-factorisation.

The \Ac{ScaLAPACK} library provides some of the linear algebra features of the \Ac{LAPACK} library for distributed matrices, most commonly leveraging the \Ac{MPI} framework for communication between nodes, which is widely available on computing clusters.
We chose to leverage the \ac{ScaLAPACK} implementation, therefore we need to take  \ac{ScaLAPACK}'s data distributing principles in consideration.
The procedure names are the same as for \Ac{LAPACK} but with a prefix \code{p}, e.g. the QR-factorisation subroutine is \code{pdgeqrf} instead of \code{dgeqrf}.
For the rest of our discussion, we will use the prefixed names.

\Ac{ScaLAPACK} asserts that matrices are block-cyclicly distributed on a 2D processor grid. This is a generalisation of cyclic and block distribution, both of which can be used as special cases.
Considering a matrix $\bA_{R \times C}$ with $R$ rows and $C$ columns, we can cut it into blocks $\ba_{r \times c}$.
The last blocks in each row or column may have fewer columns or rows.
The blocks are then distributed in a round-robin fashion amongst the processors in the $p \times q$ processors grid, wrapping around the grid until all blocks have been assigned.

For our use-case we start by considering a block distribution along the rows for a processor grid of $p \times 1$.
This entails a row block size equal to the local number of rows for each process ($\ba_{r \times C}$ and $\bb_{r \times 1}$ with $r = \lceil R / p  \rceil$).
We fill these blocks by assigning each structure (i.e. several rows of $\ba$) to a single process, thereby each atomic configuration is local on exactly one process.
The solution to $\bA \bc = \bb$ is invariant to swapping rows of $\bA$ as long as the corresponding entries in $\bb$ are swapped accordingly.
This allows us to choose the row block size freely while arriving at the same result irrespective of the assignment of atomic configurations to processes.
The column block size is unrestricted, since each row is fully assigned to a single process.

\begin{figure}
    \includegraphics[width=\columnwidth]{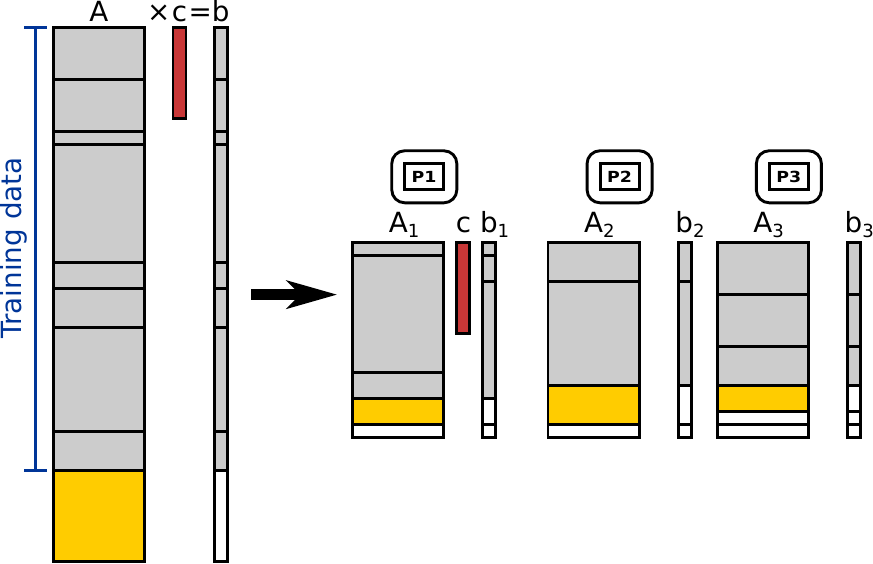}
    \centering
    \caption{Serial (left) and distributed (right) solution of $\bA \bc = \bb$. The input training data is distributed across the \Ac{MPI} processes P1, P2, P3 to balance load (but the original order of rows is preserved on each). The $\bL_{MM}^T$ matrix (yellow) in the serial implementation and its rows are distributed in the parallel implementation. Each local $\bA_i$ is filled with zero rows (white) to adjust to uniform matrix size, which is a multiple of the block size.}
    \label{fig:distributeA}
\end{figure}

Our algorithm, as described above and presented in \cref{fig:distributeA}, distributes atomic configurations and rows of $\bL_{MM}^T$ such that each local $\bA_n$ block is as equal in size as possible.
\Ac{ScaLAPACK} requires that all processes use a uniform block factor for all their blocks.
To fill the gaps left by the distribution a padding of zero rows (i.e. rows filled with zeroes) is inserted into both $\bA$ and $\bb$.
The distribution strategy and the block size settings of \ac{ScaLAPACK} should ensure that the number of padding rows are kept to a minimum to prevent afflicting memory and efficiency penalties.

%In a first run through the input, the requirements (sum of target properties $\by$, $\by'$, $\by''$) of each structure for the number of rows of $\bA$ are collected in a list. These tasks are then assigned among the available processes to level the total number of rows for each process. In a second run through the input, each process only considers their assigned structures. 

Solving the linear system via QR decomposition with \Ac{ScaLAPACK} is split into three steps.
First, $\bA$ is converted using \code{pdgeqrf} into the upper triangular matrix $\bR$ and the Householder reflectors $\bH$, which occupy the remaining lower triangular elements of $\bA$.
The latter is accompanied by an array $\btau$ of weights.
Reflectors and weights are then used by \code{pdormqr} to perform the multiplication $\bQ^T \bb$.
Finally, the linear system represented by the upper triangle matrix $\bR$ is solved by \code{pdtrtrs}, utilising the backsubstitution algorithm, to give $\bc = \bR^{-1} \bQ^T \bb$.

We note that there is a requirement in \code{pdtrtrs} that the row and column block sizes must be equal. 
Setting the column block size ($c$) to the generally much larger row block size ($r$) is formally possible, but this drastically increases the size of the working arrays the \ac{ScaLAPACK} routines require, which scale with the square of the column block size ($\propto c^2 + r c$).
Setting instead the row block size ($r$) to the column block size ($c$) implies adding additional zero rows for padding the local matrices to maintain the divisibility by the block factor and thus the assignment of the configurations to the processes.
Both of these approaches result in increased memory requirements and a deterioration of computational efficiency.

However, it is possible to exploit the fact that our distribution strategy relies on a single processor column ($q = 1$), changing the column block size does not affect the distribution of the data.
We can therefore use one column block size for the first two calls (\code{pdgeqrf}, \code{pdormqr}) and then change that value for the third call (\code{pdtrtrs}) to fulfill its requirement without decreasing the efficiency of the former calls.

Being able to control both block sizes independently revealed that a moderate column block size of about 100 is optimal for both memory usage and efficiency.
For such a setting, the row block size does not have a significant impact on parallel efficiency.

\subsection{Sparse point selection}\label{sec:sparse}

In \gapfit{}, the set of $M$ sparse points are typically determined as a subset of all descriptor values, although for low-dimensional descriptors such as bond length it is convenient to use a uniform grid.
Depending on the method, the selection of sparse points may depend on the values of descriptor vectors calculated from the entire training data set.
If the descriptors are distributed, clustering or CUR-based methods require fine-tuned communication between the processes and for simplicity, we suggest a two-step workflow.
Since the calculation of descriptor vectors -- without their gradients -- is not computationally expensive and memory storage is not a concern, sparse point selection can be performed using serial execution.

We first run \gapfit{} on a single process, optionally using \ac{OpenMP} threading, to select sparse points which are written into files, and terminating the run, which can be achieved by the command line argument \code{sparsify\_only\_no\_fit=T}.
The output is converted to input via a helper script for the subsequent run using \Ac{MPI} processes.
This step can be skipped if the sparse points file has been provided by external means or can be reused from an earlier calculation.

\subsection{Peak memory usage}

One of the pitfalls of running a formerly serial program in parallel with distributed data is that duplicate data may accumulate unnecessarily, especially if multiple processes are run on the same node, and therefore shared memory can be utilised.
For example, it is convenient to calculate the matrix $\bK_{MM}$ on each process because it only depends on the sparse points, and its size does not depend on the training set.
However, each process requires only a small part of the resulting matrix $\bL_{MM}$, and storing multiple copies of $\bK_{MM}$ adds an unnecessary constant overhead to the memory requirements.
To prevent the allocation of possibly several GB memory per process, $\bK_{MM}$ is only calculated on a single process, then converted to $\bL_{MM}$, and only the necessary rows are distributed via \code{mpi\_scatterv} calls to independent \ac{MPI} processes.

It is also important to avoid inadvertent duplication of data when converting between data structures used by different parts of the program.
This can be alleviated by performing calculations directly on the data memory as \Ac{LAPACK} does. 
For user-defined types we use pointers to the original matrices to prevent copying of data. 
Further, source data of relevant size is deallocated after final usage.
This decreases the memory overhead per \Ac{MPI} process and therefore also the peak memory requirement of the program.

\begin{figure}
    \centering
    \includegraphics[width=0.9\columnwidth]{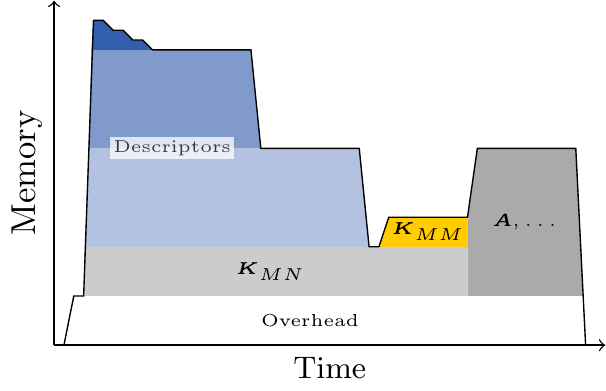}
    \caption{Schematic memory usage during \gapfit{} run over time. Descriptors (5 shown) $\bx$ and their derivatives $\bx'$ constitute the majority of the first peak. The memory associated with them is released after each processing, leading to a step-wise decline. Matrices $\bK_{MM}$ and $\bA$ and working arrays for solving the latter make up the second peak, which can be more shallow than depicted here.}
    \label{fig:memory_schema}
\end{figure}

\Cref{fig:memory_schema} shows schematically how the memory usage of \gapfit{} run changes over time.
For our program there are two parts of the execution which may lead to peak memory usage.
The main one is after allocating the descriptors, especially the derivatives $\bx'$.
After the covariance calculation of each descriptor, we deallocate the corresponding source data.
This is reflected by the step-wise decline of the memory.

The other peak manifests towards the end of a program run when the matrices $\bK_{MM}$ and then $\bA$ are assembled and the linear system is subsequently solved.
\Ac{ScaLAPACK} requires additional work arrays for some of its routines depending on the block sizes of the distributed matrices, especially the column block size.

\section{Practical Examples}

\begin{figure*}
    \centering
    \begin{subfigure}[b]{0.48\textwidth}%
        \centering%
        \includegraphics[width=\textwidth]{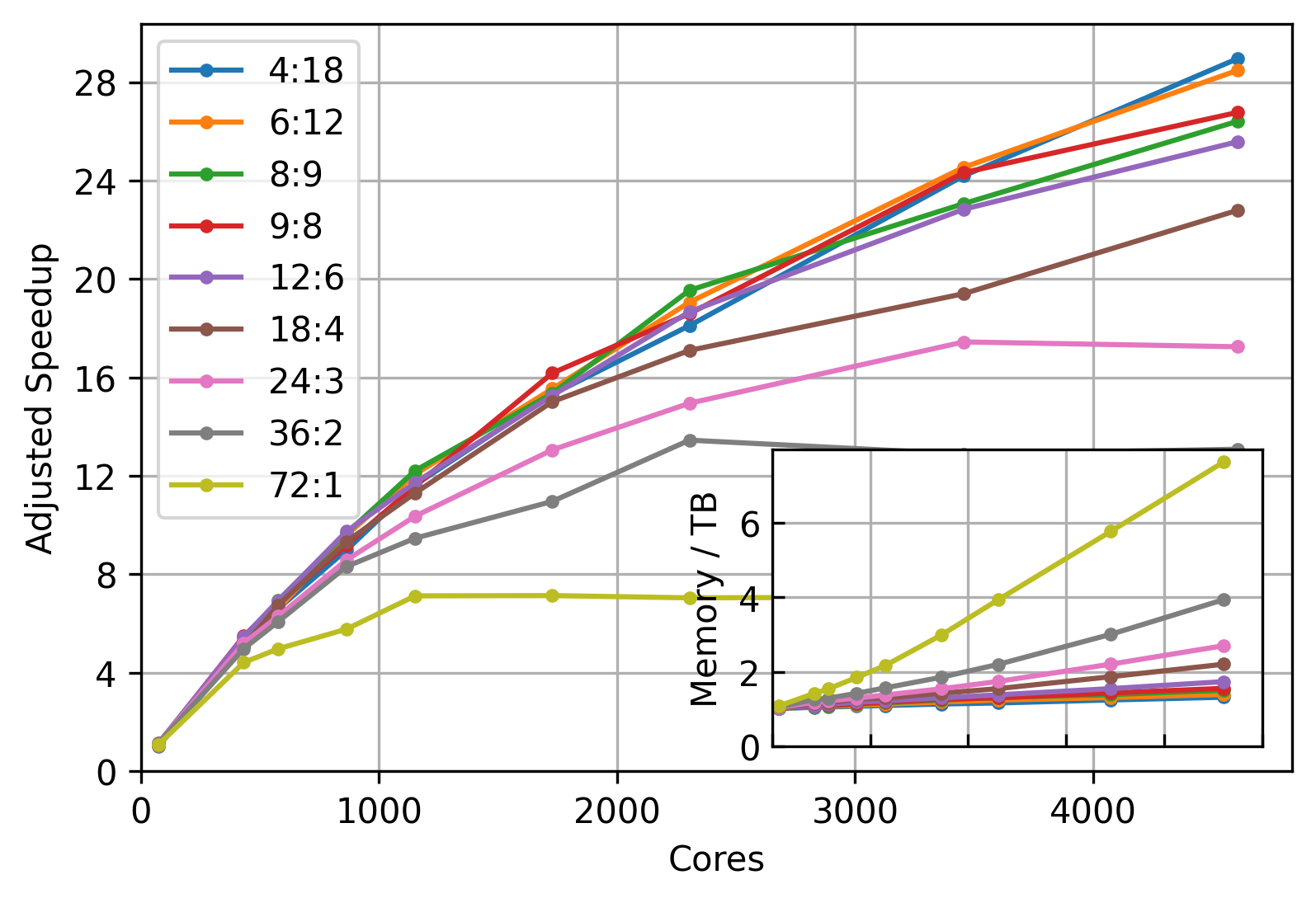}%
        \label{fig:hea4k_itime_mem}%
    \end{subfigure}
    \hfill
    \begin{subfigure}[b]{0.48\textwidth}%
        \centering%
        \includegraphics[width=\textwidth]{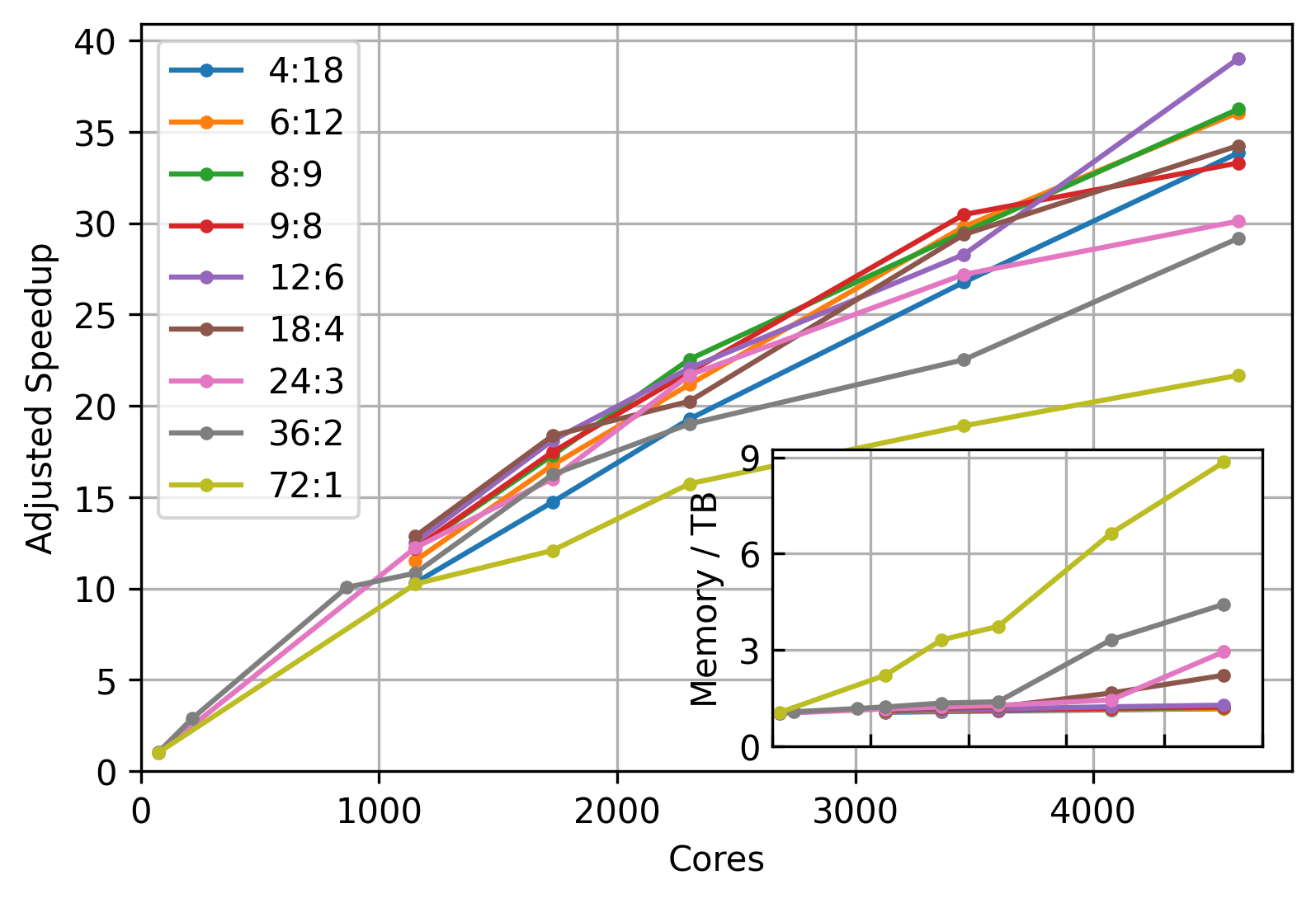}%
        \label{fig:sic10k_itime_mem}%
    \end{subfigure}
    \begin{subfigure}[b]{0.48\textwidth}%
        \centering%
        \includegraphics[width=\textwidth]{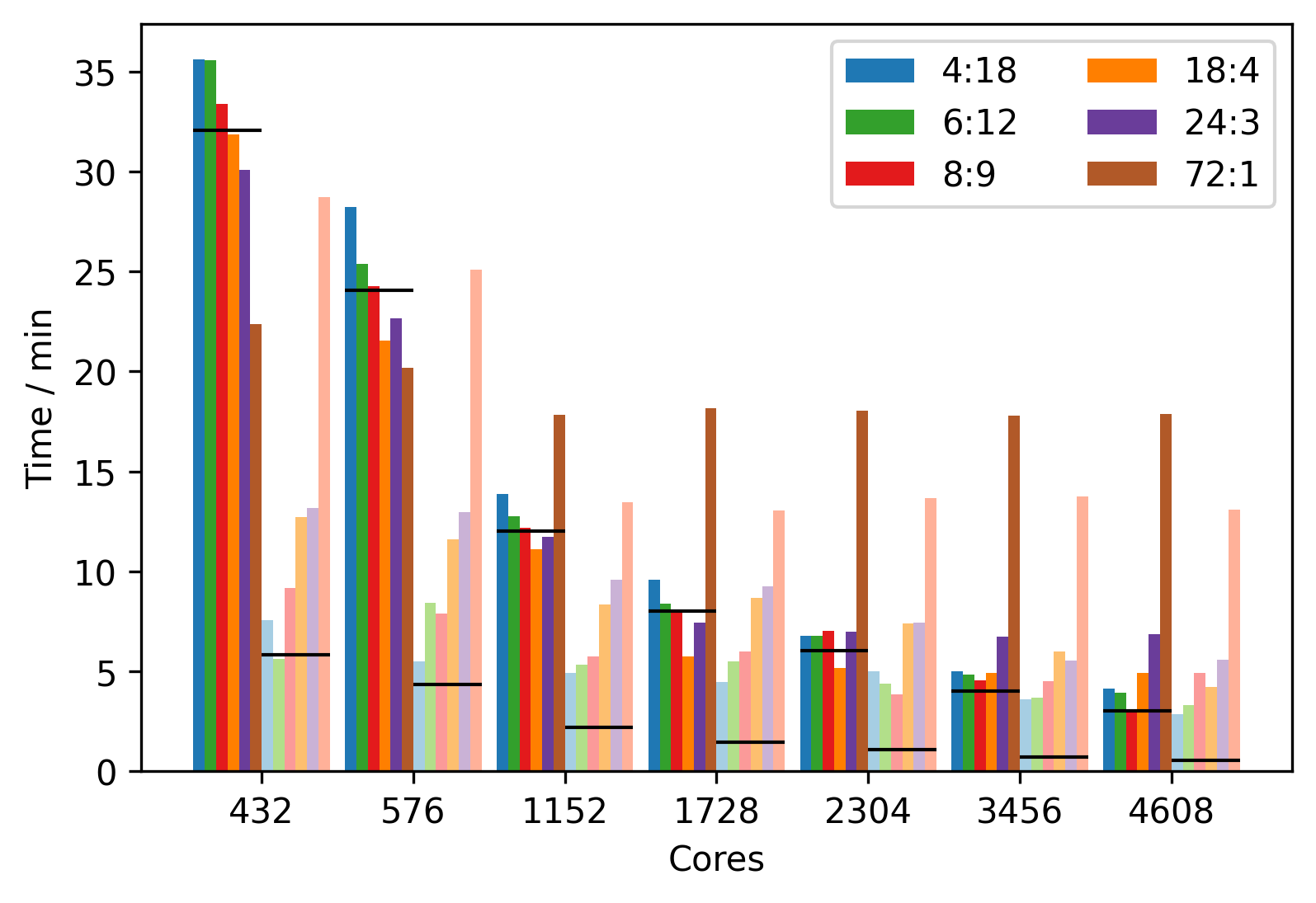}%
        \label{fig:hea4k_sub}%
    \end{subfigure}
    \hfill
    \begin{subfigure}[b]{0.48\textwidth}%
        \centering%
        \includegraphics[width=\textwidth]{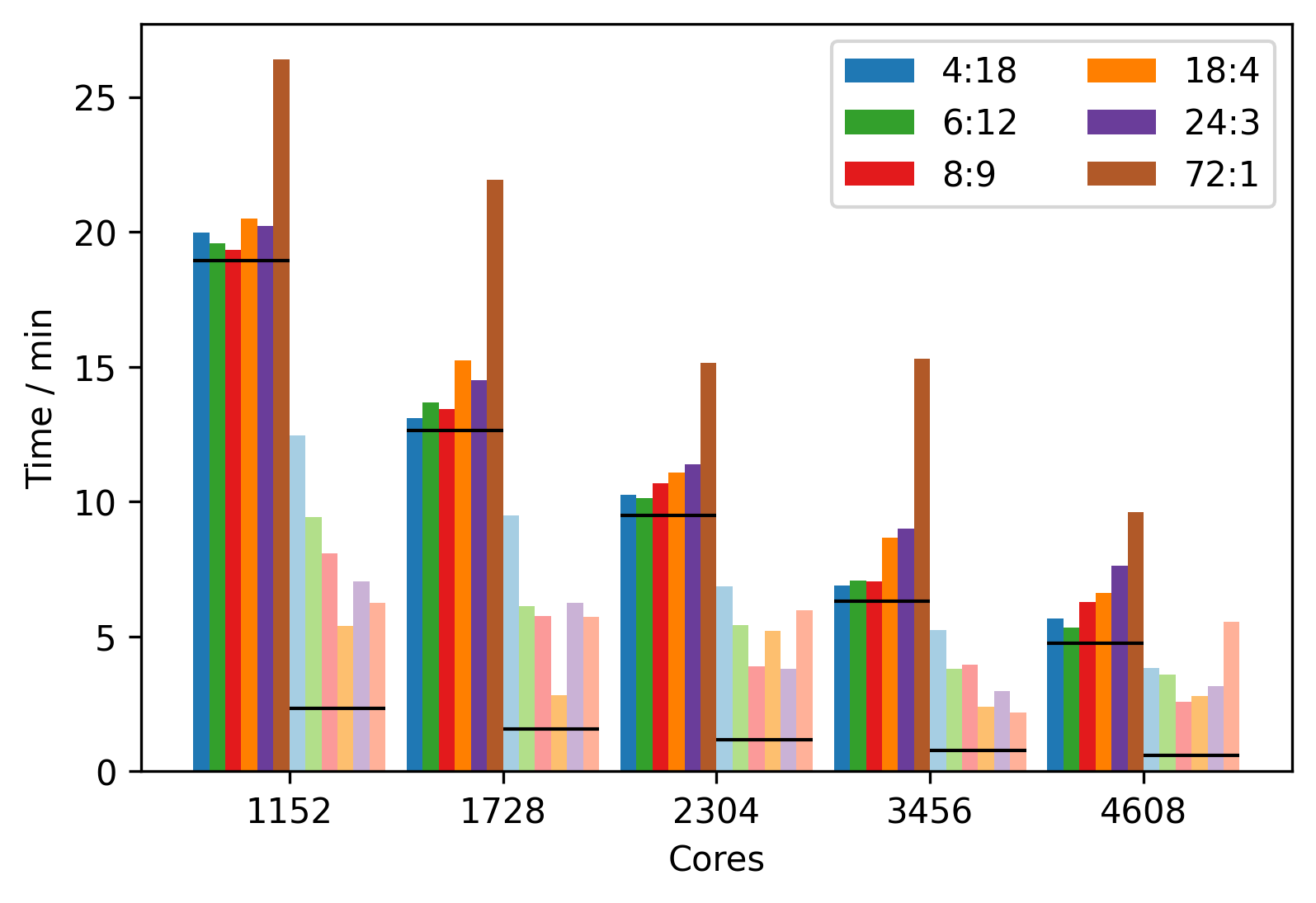}%
        \label{fig:sic10k_sub}
    \end{subfigure}
    \caption{\added{Upper: Adjusted speedup (reference time / current time) and total memory requirements (inset), and lower: Partial times for calculating the covariance matrix (dark) and solving the linear system (light); vs cores (72 per node) with different splits between \Ac{MPI} tasks per node vs \Ac{OpenMP} threads per task.
    Fitting times are shown for the HEA model (left, \num{396178} target properties) and the SiC model (right, \num{2482085} target properties), both models used \num{20300} representative (sparse) points. For visual guidance horizontal black lines indicate ideal expected partial timings, extrapolated from the partial times of the reference runs.}}
    \label{fig:examples}
\end{figure*}

Initial proof-of-principle fitting runs of a silicon dataset~\cite{Bartok.2018} showed that the fitting weights from an \Ac{MPI} run are comparable to those from serial runs when using the same representative set.
The difference can be attributed to numerical uncertainties noting that even two runs initiated with identical parameters may differ to the same magnitude.
The difference can be attributed to the fact that the order of numerical operations is non-deterministic and the floating point arithmetic is neither associative nor distributive, leading to small differences in covariance matrices in different executions.
Ill-conditioning of matrix $\bA$ amplifies the noise due to the different order of operations, leading to only a few comparable significant digits in the resulting weights.
We have therefore tested the accuracy of the predictions with the resulting potential models using the $\Delta$ metric suggested to compare \ac{DFT} packages~\cite{Lejaeghere.2016}.
We found that equivalent \ac{GAP} models differ only up to \SI{1}{\mu eV}, indicating excellent reproducibility despite the differences in the weights.

We then applied the implementation with varying proportions of \Ac{MPI} processes and \Ac{OpenMP} threads on two example training sets, consisting of an \Ac{HEA} and silicon carbide (SiC) datasets.

These calculations were performed on the Raven cluster of the Max Planck Computing and Data Facility. Each node has 72 cores and either \SI{256}{GB} or \SI{512}{GB} of RAM. For single-node calculations we used high-memory nodes with \SI{2048}{GB} of RAM. \added{The specifications are: Intel Xeon IceLake-SP processors (2.4 GHz), 72 Cores (2 NUMA domains), Mellanox HDR InfiniBand network (100 Gbit/s, pruned fat-tree topology).\cite{Raven}}

In both cases we combined two-body descriptors with \ac{SOAP} descriptors~\cite{Bartok.2013}.
We assign separate two-body descriptors for each pair of species, and separate \ac{SOAP} descriptors for each species.

\added{The upper part of} \cref{fig:examples} depicts the inverse time relation with respect to the number of cores for different \Ac{MPI} to \Ac{OpenMP} ratios (T:C), e.g. ``36:2'' means that 36 \Ac{MPI} processes were used per node (with 72 cores), each with two \Ac{OpenMP} threads.
This resembles Amdahl's law
\begin{equation}
    S(n) = t(1) / t(n) = 1 / [(1 - p) + p / n],
\end{equation}
where the speedup $S$ describes the relative time for a serial run $t(1)$ versus a parallelised one $t(n)$\added{. It} depends on the number of cores $n$ and the relative time $p$ spent in portions that benefit from multiple cores.
Our training systems were too large to be run on a single core within the maximum wall-time limit, so we used the highest time available instead for our adjusted speedup ($S^*$).
Because of this, the values are only comparable within the same training set.
Note that these timings may contain some random noise due to other calculations on the cluster (albeit not on the same nodes) and generally unpredictable behavior of interconnect, network, hardware or the operating system in general.
The insets show the total memory required for these calculations across all nodes, estimated from the average resident set size (AveRSS) as queried from the queueing system.

\added{The lower part of \cref{fig:examples} shows the absolute times for the calculation of the covariance matrix, and for preparing and solving the linear system. As a visual guide horizontal lines indicate the ideal time extrapolated from the respective partial times of the (longest) reference calculation. The remaining time to the total run time is small and not depicted for readability.}

\subsection{High-Entropy alloy}

The high-entropy MoNbTaVW alloy (\Ac{HEA}) training set~\cite{Byggmastar:2021} consists of 2329 configurations -- each containing an uneven number of atoms -- with 2329 total energies, $\num{383739}$ forces, and $\num{10110}$ virials for a total of $N = \num{396178}$ target properties.
With 20 sparse points per two-body descriptor (15) and 4000 per SOAP descriptor (5) the total number of sparse points is $M = \num{20300}$.
Thus, $\bA$ consists of $n_\bA = \num{8454503400}$ elements and occupies about \SI{67.6}{GB}.

\added{Looking at \cref{fig:examples}, the speedups for ratios between 4:18 and 12:6 do not differ much in our timings. Calculations with more MPI processes lose efficiency with increasing number of cores.}

\added{For example, omitting OpenMP (72:1) decreases the speedup gain around 500 cores, staying at constant time between 864 and 1152 cores.}
The same happens for 36:2 ratio between 1728 and 2304 cores, and for 24:3 between 2304 and 3456 cores.
This behaviour stems from the choice of our implementation: it splits the training set along the structures, which cannot be done arbitrarily for a finite set.
For these 2329 configurations, the limit is at about two structures per \Ac{MPI} process.

\added{The reference time for this set is \SI{13815}{s} (\SI{3.8}{h}), obtained from the single (high-memory) node calculation with a 4:18 split of cores. Other ratios improve only a little on top of that, the highest being 12:6, 24:3, and 36:2 with a speedup of 1.13 each. The 72:1 run does not achieve that (1.06).
Using six nodes (432 cores) increases the speedup to between 5.49 (9:8) and 4.42 (72:1), which translates to relative increases between 5.09 and 4.17 compared to a single node.
As mentioned before, the benefits dwindle fast for the 72:1 split from factor 1.31 (total 5.78) despite doubling (twelve nodes) to the constant speedup around 16 nodes of ca. 7.10.
The 36:2 ratio achieves a speedup of 13.45 at its best but decreases to about 13.0 beyond that.
The same happens, less pronounced, for 24:3, from a maximum of 17.44 at 48 cores down to 17.25.
The highest speedup in this set is 28.96 (4:18 on 64 nodes). The most efficient w.r.t. to the used resources is 18:4 on a single node.}

\added{Using more MPI processes also comes at the cost of a memory overhead, which increases approximately linearly with the number of cores. The higher the portion of \Ac{MPI} usage, the steeper this overhead is.
In fact, the graphs coincide if they are plotted against the number of \Ac{MPI} processes (not shown): in that case the slope ranges between \num{0.9} and \SI{1.5}{GB} per \Ac{MPI} process.
For 72:1 this results in \SI{2.17}{TB} on 16 nodes and \SI{7.63}{TB} on 64.
The former is comparable to the \SI{2.20}{GB} 18:4 uses on 64 nodes, since both apply 1152 \Ac{MPI} processes.}

\added{Looking at the partial timings, solving the covariance is close to the ideal performance when increasing the number of nodes.
The reference times are \SI{3.2}{h} for covariance and \SI{0.6}{h} for the QR preparation and solve from the 4:18 ratio on one node.
Calculations with a higher number of \Ac{MPI} tasks per node needs less time for a smaller number of nodes up to the point of saturation when this time stays constant.
Solving of the linear system takes less time than the covariance but does not benefit as much from an increasing number of nodes.
This is to be expected as more inter-node communication slows down the computation, while no communication is needed for the covariance.
Using all cores for \Ac{MPI} results in a severe penalty on the solving time.
The remaining runtime is negligible (up to \SI{3}{min}).}

\subsection{Silicon carbide}

The 4865 silicon carbide (SiC) systems of this training set contain 4865 energies, $\num{2448030}$ forces, $\num{29190}$ virials for a total of $N = \num{2482085}$ target properties. With 100 sparse points per two-body descriptor (3) and $\num{10000}$ per SOAP descriptor (2) the total number of sparse points is $M = \num{20300}$.
Thus, $\bA$ consist of $n_\bA = \num{50798415500}$ elements and occupies \SI{406.4}{GB}.

Due to the much larger training set, not all node configurations from the \Ac{HEA} set were viable, especially for lower node numbers.
\added{The reference time is \SI{20472}{s} (\SI{5.7}{h}) for a 24:3 run on a single node.
The trend of an equal split between MPI and OpenMP being the preferred choice also holds for this set but the processes are not saturated as rapidly as in the case of the \Ac{HEA} systems.}
This effect may start between 3456 and 4608 cores for the full \Ac{MPI} run (72:1), which is later than in the \Ac{HEA} set even taking the structure numbers into account because of the proportionally lower number of small structures in this larger set of only two species.
\added{The highest speedup of 38.99 is achieved by 12:6 on 64 nodes, similarly 18:4 split reaches the best value of 12.80 on 16 nodes whereas 36:2 gives 10.84 and 4:18 only 10.31.}

\added{Our method of querying the queueing system underestimates the average memory usage for some of the calculations. The general trend is, however, relatable to the HEA case, as shown by a similar values for high core numbers. The memory overhead per \Ac{MPI} process is nominally between \num{0.3} and \SI{1.8}{GB}. On 16 nodes the memory usage for 72:1 is given as \SI{2.22}{TB}, on 64 nodes as \SI{8.85}{TB}.}

\added{For the partial timings we have similar trends as in the HEA case.
The reference times are \SI{5.1}{h} for covariance and \SI{0.6}{h} for the QR preparation and solve from the 24:3 ratio on one node.
Strikingly, the 72:1 covariance times are much higher than for other splits.
The other times are quite similar between different ratios, in agreement with the total speedup.
The remaining runtime takes up to \SI{1}{min}.}

\section{Conclusion and Outlook}

The recent addition of \Ac{MPI} parallelisation to our program \gapfit{} by using the \Ac{ScaLAPACK} library makes it possible to split the training data evenly into batches to be processed independently up to the solving of the linear system.
It alleviates the need for high-memory nodes so commodity HPC nodes may be used for arbitrarily large fitting problems. \replaced{The efficiency of the MPI parallelisation is comparable to that of OpenMP for all parallelisation strategies except fully MPI, effectively extending the previous efficiency to multiple nodes.}{while computation time has been significantly reduced the due to the pleasingly parallel algorithm.}
Thus larger training sets do not impede the computation and more sparse points can be used, increasing the accuracy of the model.

We showed the time scaling and memory requirements for varying proportions of \Ac{MPI} processes vs \Ac{OpenMP} threads in two example training sets, consisting of an high-entropy alloy (\Ac{HEA}) and silicon carbide (SiC).
It is generally advisable to use \replaced{at least some cores on each node for \Ac{MPI} processes}{most of the processors for}, so even on a single node benefits from this new feature.
It is especially effective for larger training sets while the sparse points are covered by the \Ac{OpenMP} threads.
One should keep the total number of \Ac{MPI} processes below some fraction of the total number of structures, e.g. 0.5 so that an even distribution is still possible.
\added{Analysing the partial times of calculating the covariance matrix and solving the linear system confirmed that the benefit stems from the almost perfectly parallel covariance part due to the splitting of configurations. The inter-node communication of the \Ac{ScaLAPACK} routines prevents the solving part to benefit as much from a larger number of nodes.}

The memory overhead due to the parallelisation has been reduced but is still significant.
Depending on the available memory resources, a higher share of \Ac{OpenMP} threads is preferable\added{, coinciding with the conclusion from the efficiency evaluation}. \replaced{An even smaller memory footprint may be achieved in a further development, albeit the impact is not as severe since the recommended number of MPI processes is moderate.}{We are confident that an even smaller memory footprint may be achieved in a further development.}

The highest impact on both \Ac{MPI} and non-\Ac{MPI} memory requirements would be to fully restructure the descriptor processing loop so that each descriptor is processed fully before the next one.

In practical tests we have seen that the parallel \code{gap\_fit} code can decrease the time required to fit potentials from days to \added{tens of} minutes.
We anticipate this will be an important step to enable potential fitting to be embedded within other higher-level workflows such as active learning~\cite{Li2015,Vandermause2020}, committee models~\cite{Imbalzano2021} as well as enabling model hyperparameters to be tuned or optimised, known to be important for improved uncertainty quantification~\cite{Vandermause2020}.

\begin{acknowledgments}

We thank Harry Tunstall for providing the SiC dataset and early testing as well as G\'abor Cs\'anyi and Miguel Caro for useful discussions.
% anyone else
This work was financially supported by the NOMAD Centre of Excellence (European Commission grant agreement ID 951786) and the Leverhulme Trust Research Project Grant (RPG-2017-191).
ABP acknowledges support from the CASTEP-USER project, funded by the Engineering and Physical Sciences Research Council under the grant agreement EP/W030438/1. 
We acknowledge computational resources provided by the Max Planck Computing and Data Facility provided through the NOMAD CoE, the Scientific Computing Research Technology Platform of the University of Warwick, the EPSRC-funded HPC Midlands+ consortium (EP/T022108/1) and ARCHER2 (https://www.archer2.ac.uk/) via the UK Car-Parinello consortium (EP/P022065/1). We thank the technical staff at each of these HPC centres for their support.

\added{For the purpose of open access, the author has applied a Creative Commons Attribution (CC BY) licence to any Author Accepted Manuscript version arising from this submission.}

\end{acknowledgments}

\bibliographystyle{apsrev4-2}
%\bibliography{inc/gap_fit}% Produces the bibliography via BibTeX.

\appendix
\onecolumngrid

\section{\gapfit{} options and timings}

{\setlength{\parskip}{1em}
\raggedright

The following input is split into separate entries here for readability. It needs to be expanded for use.

For the \ac{HEA} set:

{\ttfamily
at\_file=data.xyz gap=GAP default\_sigma=\{0.005 0.2 0.05 0.0\} energy\_parameter\_name=dft\_energy force\_parameter\_name=dft\_forces sparse\_jitter=1.0E-8 gp\_file=gap.xml rnd\_seed=999
}

{\ttfamily
GAP=\{\{SUBGAP1\}:\{SUBGAP2\}\}
}

{\ttfamily
SUBGAP1=distance\_2b cutoff=4.5 delta=10.0 covariance\_type=ard\_se theta\_uniform=0.75 n\_sparse=20 sparse\_method=uniform
}

{\ttfamily
SUBGAP2=soap l\_max=4 n\_max=8 atom\_sigma=0.5 zeta=2 cutoff=3.5 cutoff\_transition\_width=0.5 central\_weight=1.0 n\_sparse=4000 delta=0.1 covariance\_type=dot\_product sparse\_method=cur\_points
}

For the SiC set: {\ttfamily rnd\_seed=84616238}, {\ttfamily n\_sparse=100} and {\ttfamily n\_sparse=10000}, respectively.

}

\include{tab/HEA}
\include{tab/SiC}

\end{document}

%% file: inc/preamble.tex
\usepackage{graphicx}
\usepackage{dcolumn}
\usepackage{bm}
\usepackage{xfrac}
\usepackage{xspace}
\usepackage{siunitx}
\usepackage{hyperref} % load before cleveref
\usepackage[capitalise]{cleveref}
\usepackage{booktabs}
\usepackage{caption}
\usepackage{subcaption}
\usepackage{float}
\usepackage{placeins}
\usepackage{enumitem}

\newlist{commalist}{description*}{1}
\setlist[commalist]{
    itemjoin={{,}},
    itemjoin*={{, and}},
    afterlabel=\unskip{{~}}
}

\newlist{arglist}{description}{1}
\setlist[arglist]{
  itemsep=0ex,
  leftmargin=1em,
  labelsep=1em,
  align=left,
  font=\ttfamily\bfseries,
  before={}
}

\usepackage[final,commentmarkup=footnote]{changes}

\usepackage{acro}

\DeclareAcronym{GAP}{short=GAP, long=Gaussian Approximation Potential}
\DeclareAcronym{GP}{short=GP, long=Gaussian Process}
\DeclareAcronym{SOAP}{short=SOAP, long=Smooth Overlap of Atomic Positions}
\DeclareAcronym{DFT}{short=DFT, long=Density Functional Theory}
\DeclareAcronym{RMSE}{short=RMSE, long=root mean squared error}
\DeclareAcronym{QUIP}{short=QUIP, long=Quantum Mechanics and Interatomic Potentials}
\DeclareAcronym{MPI}{short=MPI, long=Message Passing Interface}
\DeclareAcronym{OpenMP}{short=OpenMP, long=Open Multiprocessing}
\DeclareAcronym{LAPACK}{short=LAPACK, long=Linear Algebra Package}
\DeclareAcronym{ScaLAPACK}{short=ScaLAPACK, long=Scalable \Ac{LAPACK}}
\DeclareAcronym{HEA}{short=HEA, long=High-Entropy Alloy}
\DeclareAcronym{MLIP}{short=MLIP, long=machine learning interatomic potential}
\DeclareAcronym{SNAP}{short=SNAP, long=Spectral Neighbour Analysis Potential}
\DeclareAcronym{LML}{short=LML, long=Linear Machine Learning}
\DeclareAcronym{ACE}{short=ACE, long=Atomic Cluster Expansion}

\input{inc/symbols}

\graphicspath{ {img/} }

%% file: inc/symbols.tex
\usepackage{dsfont}
\usepackage{bm,upgreek}

\newcommand{\gapfit}{\texttt{gap\_fit}\xspace{}}
\newcommand{\abinitio}{\textit{ab initio\xspace}}
\newcommand{\etal}{\textit{et al\xspace}}

\newcommand{\kc}{c}

\newcommand{\ba}{\mathbf{a}}
\newcommand{\bb}{\mathbf{b}}
\newcommand{\bA}{\mathbf{A}}

\newcommand{\bc}{\mathbf{\kc}}
\newcommand{\bH}{\mathbf{H}}
\newcommand{\bI}{\mathbf{I}}

\newcommand{\bK}{\mathbf{K}}
\newcommand{\bL}{\mathbf{L}}

\newcommand{\bQ}{\mathbf{Q}}
\newcommand{\bR}{\mathbf{R}}
\newcommand{\bx}{\mathbf{x}}
\newcommand{\by}{\mathbf{y}}

\newcommand{\btau}{\mathbf{\tau}}

\newcommand{\bSigma}{\boldsymbol{\Sigma}}

\newcommand{\code}[1]{\texttt{#1}}

%% file: inc/preface.tex
%\preprint{APS/123-QED}

\title{Massively Parallel Fitting of Gaussian Approximation Potentials}

\author{Sascha Klawohn}
\affiliation{Warwick Centre for Predictive Modelling, School of Engineering, University of Warwick, Coventry CV4 7AL, United Kingdom}

\author{James R. Kermode}
\affiliation{Warwick Centre for Predictive Modelling, School of Engineering, University of Warwick, Coventry CV4 7AL, United Kingdom}

\author{Albert P. Bart\'ok}
\affiliation{Department of Physics, University of Warwick, Coventry CV4 7AL, United Kingdom}
\affiliation{Warwick Centre for Predictive Modelling, School of Engineering, University of Warwick, Coventry CV4 7AL, United Kingdom}
\email{apbartok@gmail.com}

\date{\today}

\begin{abstract}
We present a data-parallel software package for fitting Gaussian Approximation Potentials (GAPs) on multiple nodes using the ScaLAPACK library with MPI and OpenMP. Until now the maximum training set size for GAP models has been limited by the available memory on a single compute node. In our new implementation, descriptor evaluation is carried out in parallel with no communication requirement. The subsequent linear solve required to determine the model coefficients is parallelised with ScaLAPACK. Our approach scales to thousands of cores, lifting the memory limitation and also delivering substantial speedups. This development expands the applicability of the GAP approach to more complex systems as well as opening up opportunities for efficiently embedding GAP model fitting within higher-level workflows such as committee models or hyperparameter optimisation.
\end{abstract}

%% file: tab/HEA.tex
\begin{table}[hb]
    \scriptsize
    \renewcommand{\arraystretch}{0.70}
    \begin{tabular}{rrrrrrrr}
        \toprule
         $N$ &  $T$ & $C$ & $t_\text{Cov}$ & $t_\text{QR}$ & $t$ &  $S^*$   &   Mem \\
        \midrule
          1 &   4 &  18 & 11545.54 & 2090.22 &  13815 &  1.00 & 1.02 \\
          1 &   6 &  12 & 11461.97 & 1450.42 &  13052 &  1.06 & 1.02 \\
          1 &   8 &   9 & 11580.61 &  998.30 &  12699 &  1.09 & 1.03 \\
          1 &   9 &   8 & 11324.90 & 1316.42 &  12757 &  1.08 & 1.03 \\
          1 &  12 &   6 & 11515.25 &  756.68 &  12376 &  1.12 & 1.03 \\
          1 &  18 &   4 & 11042.48 & 1072.07 &  12208 &  1.13 & 1.04 \\
          1 &  24 &   3 & 11613.63 &  562.52 &  12262 &  1.13 & 1.04 \\
          1 &  36 &   2 & 11571.76 &  596.12 &  12248 &  1.13 & 1.05 \\
          1 &  72 &   1 & 11128.19 & 1798.22 &  13002 &  1.06 & 1.09 \\
          6 &   4 &  18 &  2136.17 &  454.57 &   2706 &  5.11 & 1.05 \\
          6 &   6 &  12 &  2134.75 &  338.10 &   2540 &  5.44 & 1.06 \\
          6 &   8 &   9 &  2003.19 &  549.59 &   2618 &  5.28 & 1.07 \\
          6 &   9 &   8 &  2019.54 &  431.06 &   2516 &  5.49 & 1.08 \\
          6 &  12 &   6 &  1955.56 &  512.57 &   2529 &  5.46 & 1.10 \\
          6 &  18 &   4 &  1910.88 &  762.27 &   2734 &  5.05 & 1.14 \\
          6 &  24 &   3 &  1804.70 &  789.55 &   2653 &  5.21 & 1.18 \\
          6 &  36 &   2 &  1753.38 &  963.17 &   2775 &  4.98 & 1.26 \\
          6 &  72 &   1 &  1341.73 & 1724.40 &   3126 &  4.42 & 1.42 \\
          8 &   4 &  18 &  1692.67 &  330.79 &   2093 &  6.60 & 1.06 \\
          8 &   6 &  12 &  1522.45 &  506.49 &   2098 &  6.58 & 1.08 \\
          8 &   8 &   9 &  1457.35 &  472.29 &   1993 &  6.93 & 1.09 \\
          8 &   9 &   8 &  1458.57 &  506.79 &   2028 &  6.81 & 1.10 \\
          8 &  12 &   6 &  1423.03 &  511.80 &   1995 &  6.92 & 1.13 \\
          8 &  18 &   4 &  1293.30 &  696.93 &   2048 &  6.75 & 1.18 \\
          8 &  24 &   3 &  1360.67 &  777.20 &   2196 &  6.29 & 1.22 \\
          8 &  36 &   2 &  1287.31 &  929.63 &   2274 &  6.08 & 1.30 \\
          8 &  72 &   1 &  1211.06 & 1506.51 &   2779 &  4.97 & 1.54 \\
         12 &   4 &  18 &  1094.36 &  377.24 &   1536 &  8.99 & 1.08 \\
         12 &   6 &  12 &  1043.46 &  331.62 &   1440 &  9.59 & 1.10 \\
         12 &   8 &   9 &  1004.06 &  366.18 &   1428 &  9.67 & 1.13 \\
         12 &   9 &   8 &   974.53 &  467.68 &   1501 &  9.20 & 1.15 \\
         12 &  12 &   6 &   915.00 &  444.80 &   1417 &  9.75 & 1.19 \\
         12 &  18 &   4 &   876.25 &  546.52 &   1480 &  9.33 & 1.26 \\
         12 &  24 &   3 &   891.94 &  660.54 &   1609 &  8.59 & 1.30 \\
         12 &  36 &   2 &   696.66 &  908.88 &   1661 &  8.32 & 1.42 \\
         12 &  72 &   1 &  1076.56 & 1255.26 &   2392 &  5.78 & 1.85 \\
         16 &   4 &  18 &   831.71 &  293.98 &   1187 & 11.64 & 1.10 \\
         16 &   6 &  12 &   766.59 &  320.15 &   1145 & 12.07 & 1.13 \\
         16 &   8 &   9 &   731.07 &  343.83 &   1131 & 12.21 & 1.17 \\
         16 &   9 &   8 &   695.14 &  438.43 &   1192 & 11.59 & 1.19 \\
         16 &  12 &   6 &   696.03 &  425.87 &   1178 & 11.73 & 1.22 \\
        \bottomrule
    \end{tabular}
    \hspace{1em}
    \begin{tabular}{rrrrrrrr}
        \toprule
         $N$ &  $T$ & $C$ & $t_\text{Cov}$ & $t_\text{QR}$ & $t$ &  $S^*$   &   Mem \\
        \midrule
         16 &  18 &   4 &   666.86 &  499.93 &   1222 & 11.31 & 1.30 \\
         16 &  24 &   3 &   702.33 &  574.44 &   1333 & 10.36 & 1.38 \\
         16 &  36 &   2 &   619.84 &  782.44 &   1459 &  9.47 & 1.57 \\
         16 &  72 &   1 &  1070.94 &  807.78 &   1939 &  7.12 & 2.17 \\
         24 &   4 &  18 &   575.69 &  268.59 &    902 & 15.32 & 1.14 \\
         24 &   6 &  12 &   504.06 &  329.16 &    889 & 15.54 & 1.19 \\
         24 &   8 &   9 &   485.88 &  358.45 &    900 & 15.35 & 1.24 \\
         24 &   9 &   8 &   478.54 &  319.92 &    854 & 16.18 & 1.27 \\
         24 &  12 &   6 &   469.10 &  381.74 &    905 & 15.27 & 1.30 \\
         24 &  18 &   4 &   344.33 &  519.75 &    920 & 15.02 & 1.43 \\
         24 &  24 &   3 &   445.16 &  555.32 &   1058 & 13.06 & 1.55 \\
         24 &  36 &   2 &   555.51 &  647.66 &   1260 & 10.96 & 1.86 \\
         24 &  72 &   1 &  1090.61 &  783.59 &   1936 &  7.14 & 2.99 \\
         32 &   4 &  18 &   407.57 &  299.81 &    763 & 18.11 & 1.17 \\
         32 &   6 &  12 &   407.41 &  262.24 &    725 & 19.06 & 1.24 \\
         32 &   8 &   9 &   421.34 &  231.88 &    707 & 19.54 & 1.31 \\
         32 &   9 &   8 &   374.51 &  309.80 &    743 & 18.59 & 1.30 \\
         32 &  12 &   6 &   373.78 &  311.40 &    740 & 18.67 & 1.38 \\
         32 &  18 &   4 &   310.40 &  443.29 &    808 & 17.10 & 1.55 \\
         32 &  24 &   3 &   420.02 &  444.83 &    924 & 14.95 & 1.74 \\
         32 &  36 &   2 &   557.74 &  414.69 &   1027 & 13.45 & 2.20 \\
         32 &  72 &   1 &  1081.61 &  820.29 &   1961 &  7.04 & 3.93 \\
         48 &   4 &  18 &   298.81 &  216.20 &    571 & 24.19 & 1.25 \\
         48 &   6 &  12 &   289.29 &  219.85 &    563 & 24.54 & 1.31 \\
         48 &   8 &   9 &   273.59 &  270.72 &    599 & 23.06 & 1.39 \\
         48 &   9 &   8 &   213.21 &  299.95 &    568 & 24.32 & 1.43 \\
         48 &  12 &   6 &   228.32 &  322.18 &    605 & 22.83 & 1.55 \\
         48 &  18 &   4 &   295.86 &  359.62 &    712 & 19.40 & 1.87 \\
         48 &  24 &   3 &   403.50 &  332.40 &    792 & 17.44 & 2.21 \\
         48 &  36 &   2 &   548.22 &  467.59 &   1072 & 12.89 & 3.01 \\
         48 &  72 &   1 &  1067.96 &  825.62 &   1952 &  7.08 & 5.77 \\
         64 &   4 &  18 &   248.17 &  171.94 &    477 & 28.96 & 1.32 \\
         64 &   6 &  12 &   234.53 &  197.51 &    485 & 28.48 & 1.39 \\
         64 &   8 &   9 &   175.12 &  293.97 &    523 & 26.41 & 1.51 \\
         64 &   9 &   8 &   177.17 &  283.84 &    516 & 26.77 & 1.56 \\
         64 &  12 &   6 &   230.36 &  255.08 &    540 & 25.58 & 1.74 \\
         64 &  18 &   4 &   296.01 &  254.14 &    606 & 22.80 & 2.20 \\
         64 &  24 &   3 &   411.00 &  335.30 &    801 & 17.25 & 2.70 \\
         64 &  36 &   2 &   545.20 &  454.38 &   1056 & 13.08 & 3.94 \\
         64 &  72 &   1 &  1072.27 &  784.67 &   1916 &  7.21 & 7.63 \\
        \bottomrule
    \end{tabular}
    \caption{\added{Time $t$ (in s), partial times $t_\text{cov}$ for covariance matrix and $t_\text{QR}$ for the QR solve, adjusted speedup $S^*$, and estimated memory (in TB) in the HEA training set. N: nodes, T: MPI tasks per node, C: OpenMP threads per task.}}
\end{table}

\begin{table}[hb]
    \renewcommand{\arraystretch}{0.75}
    \begin{tabular}{rrrrrrrrrr}
        \toprule & \multicolumn{9}{c}{$T$} \\
        \cmidrule{2-10}
        t &  4  &    6  &    8  &    9  &    12 &    18 &    24 &    36 &   72 \\
        \midrule
        1  &  1.00 &  1.06 &  1.09 &  1.08 &  1.12 &  1.13 &  1.13 &  1.13 & 1.06 \\
        6  &  5.11 &  5.44 &  5.28 &  5.49 &  5.46 &  5.05 &  5.21 &  4.98 & 4.42 \\
        8  &  6.60 &  6.58 &  6.93 &  6.81 &  6.92 &  6.75 &  6.29 &  6.08 & 4.97 \\
        12 &  8.99 &  9.59 &  9.67 &  9.20 &  9.75 &  9.33 &  8.59 &  8.32 & 5.78 \\
        16 & 11.64 & 12.07 & 12.21 & 11.59 & 11.73 & 11.31 & 10.36 &  9.47 & 7.12 \\
        24 & 15.32 & 15.54 & 15.35 & 16.18 & 15.27 & 15.02 & 13.06 & 10.96 & 7.14 \\
        32 & 18.11 & 19.06 & 19.54 & 18.59 & 18.67 & 17.10 & 14.95 & 13.45 & 7.04 \\
        48 & 24.19 & 24.54 & 23.06 & 24.32 & 22.83 & 19.40 & 17.44 & 12.89 & 7.08 \\
        64 & 28.96 & 28.48 & 26.41 & 26.77 & 25.58 & 22.80 & 17.25 & 13.08 & 7.21 \\
        \bottomrule
    \end{tabular}
    \caption{\added{Pivot table of the adjusted speedups for the HEA training set.}}
\end{table}

%% file: tab/SiC.tex
\begin{table}[h!t]
    \renewcommand{\arraystretch}{0.70}
    \begin{tabular}{rrrrrrrr}
        \toprule
         $N$ &  $T$ & $C$ & $t_\text{Cov}$ & $t_\text{QR}$ & $t$ &  $S^*$   &   Mem \\
        \midrule
          1 &  24 &   3 & 18185.23 & 2232.89 &  20472 &  1.00 & 1.03 \\
          1 &  36 &   2 & 18124.19 & 1736.61 &  19911 &  1.03 & 1.04 \\
          1 &  72 &   1 & 18551.53 & 1667.12 &  20265 &  1.01 & 1.05 \\
          3 &  36 &   2 &  6247.50 &  780.98 &   7067 &  2.90 & 1.08 \\
         12 &  36 &   2 &  1796.59 &  200.76 &   2035 & 10.06 & 1.18 \\
         16 &   4 &  18 &  1198.94 &  746.59 &   1986 & 10.31 & 1.06 \\
         16 &   6 &  12 &  1175.45 &  565.38 &   1776 & 11.53 & 1.08 \\
         16 &   8 &   9 &  1160.58 &  485.56 &   1681 & 12.18 & 1.10 \\
         16 &   9 &   8 &  1198.35 &  442.01 &   1677 & 12.21 & 1.11 \\
         16 &  12 &   6 &  1162.79 &  437.15 &   1634 & 12.53 & 1.13 \\
         16 &  18 &   4 &  1229.67 &  322.71 &   1592 & 12.86 & 1.14 \\
         16 &  24 &   3 &  1212.87 &  422.35 &   1674 & 12.23 & 1.17 \\
         16 &  36 &   2 &  1317.43 &  532.71 &   1889 & 10.84 & 1.23 \\
         16 &  72 &   1 &  1583.19 &  374.91 &   1998 & 10.25 & 2.22 \\
         24 &   4 &  18 &   785.98 &  568.33 &   1390 & 14.73 & 1.09 \\
         24 &   6 &  12 &   820.59 &  367.30 &   1222 & 16.75 & 1.11 \\
         24 &   8 &   9 &   806.53 &  345.81 &   1184 & 17.29 & 1.14 \\
         24 &   9 &   8 &   837.33 &  300.58 &   1172 & 17.47 & 1.15 \\
         24 &  12 &   6 &   824.66 &  274.41 &   1131 & 18.10 & 1.14 \\
         24 &  18 &   4 &   913.24 &  168.86 &   1115 & 18.36 & 1.18 \\
         24 &  24 &   3 &   869.64 &  375.20 &   1279 & 16.01 & 1.23 \\
         24 &  36 &   2 &  1058.73 &  168.08 &   1261 & 16.23 & 1.35 \\
         24 &  72 &   1 &  1315.80 &  342.81 &   1695 & 12.08 & 3.32 \\ 
         32 &   4 &  18 &   614.76 &  411.59 &   1062 & 19.28 & 1.10 \\
         32 &   6 &  12 &   607.02 &  325.76 &    967 & 21.17 & 1.14 \\
        \bottomrule
    \end{tabular}
    \hspace{1em}
    \begin{tabular}{rrrrrrrr}
        \toprule
         $N$ &  $T$ & $C$ & $t_\text{Cov}$ & $t_\text{QR}$ & $t$ &  $S^*$   &   Mem \\
        \midrule
         32 &   8 &   9 &   640.56 &  232.78 &    908 & 22.55 & 1.17 \\
         32 &   9 &   8 &   627.46 &  275.51 &    937 & 21.85 & 1.15 \\
         32 &  12 &   6 &   602.28 &  291.93 &    928 & 22.06 & 1.17 \\
         32 &  18 &   4 &   665.32 &  312.82 &   1011 & 20.25 & 1.23 \\
         32 &  24 &   3 &   682.36 &  228.29 &    944 & 21.69 & 1.28 \\
         32 &  36 &   2 &   802.91 &  241.16 &   1077 & 19.01 & 1.39 \\
         32 &  72 &   1 &   908.52 &  358.63 &   1301 & 15.74 & 3.74 \\
         48 &   4 &  18 &   413.15 &  314.09 &    765 & 26.76 & 1.14 \\
         48 &   6 &  12 &   424.74 &  228.41 &    687 & 29.80 & 1.15 \\
         48 &   8 &   9 &   423.08 &  237.51 &    694 & 29.50 & 1.18 \\
         48 &   9 &   8 &   471.31 &  166.77 &    672 & 30.46 & 1.19 \\
         48 &  12 &   6 &   446.73 &  244.14 &    724 & 28.28 & 1.24 \\
         48 &  18 &   4 &   519.34 &  142.77 &    697 & 29.37 & 1.66 \\
         48 &  24 &   3 &   540.43 &  178.59 &    753 & 27.19 & 1.45 \\
         48 &  36 &   2 &   653.71 &  220.68 &    909 & 22.52 & 3.32 \\
         48 &  72 &   1 &   917.48 &  129.88 &   1083 & 18.90 & 6.63 \\
         64 &   4 &  18 &   340.06 &  230.36 &    605 & 33.84 & 1.18 \\
         64 &   6 &  12 &   319.72 &  214.31 &    568 & 36.04 & 1.18 \\
         64 &   8 &   9 &   376.70 &  155.10 &    565 & 36.23 & 1.22 \\
         64 &   9 &   8 &   353.46 &  228.42 &    615 & 33.29 & 1.24 \\
         64 &  12 &   6 &   350.17 &  141.57 &    525 & 38.99 & 1.29 \\
         64 &  18 &   4 &   395.85 &  167.28 &    598 & 34.23 & 2.22 \\
         64 &  24 &   3 &   456.60 &  189.24 &    680 & 30.11 & 2.95 \\
         64 &  36 &   2 &   447.39 &  217.53 &    702 & 29.16 & 4.43 \\
         64 &  72 &   1 &   576.34 &  331.77 &    945 & 21.66 & 8.85 \\
        \bottomrule
    \end{tabular}
    \caption{\added{Time $t$ (in s), partial times $t_\text{cov}$ for covariance matrix and $t_\text{QR}$ for the QR solve, adjusted speedup $S^*$, and estimated memory (in TB) in the SiC training set. N: nodes, T: MPI tasks per node, C: OpenMP threads per task.}}
\end{table}
\setlength{\tabcolsep}{4pt}
\begin{table}[hb]
    \renewcommand{\arraystretch}{0.75}
    \begin{tabular}{rrrrrrrrrr}
        \toprule
          & \multicolumn{9}{c}{$T$} \\
        \cmidrule{2-10}
        $N$ &   4  &    6  &    8  &    9  &    12 &    18 &    24 &    36 &    72 \\
        \midrule
         1 &       &       &       &       &       &       &  1.00 &  1.03 &  1.01 \\
         3 &       &       &       &       &       &       &       &  2.90 &       \\
        12 &       &       &       &       &       &       &       & 10.06 &       \\
        16 & 10.31 & 11.53 & 12.18 & 12.21 & 12.53 & 12.86 & 12.23 & 10.84 & 10.25 \\
        24 & 14.73 & 16.75 & 17.29 & 17.47 & 18.10 & 18.36 & 16.01 & 16.23 & 12.08 \\
        32 & 19.28 & 21.17 & 22.55 & 21.85 & 22.06 & 20.25 & 21.69 & 19.01 & 15.74 \\
        48 & 26.76 & 29.80 & 29.50 & 30.46 & 28.28 & 29.37 & 27.19 & 22.52 & 18.90 \\
        64 & 33.84 & 36.04 & 36.23 & 33.29 & 38.99 & 34.23 & 30.11 & 29.16 & 21.66 \\
        \bottomrule
    \end{tabular}
    \caption{\added{Pivot table of the adjusted speedups for the SiC training set. N: nodes, T: MPI processes per node.}}
\end{table}